\documentclass{emulateapj}
\usepackage{graphicx}
\usepackage{amsmath}
\shorttitle{Differential Morphologies of $1.5<z<3$ SFGs}
\shortauthors{Bond et al.}

\begin{document}

\title{DIFFERENTIAL MORPHOLOGY BETWEEN REST-FRAME OPTICAL AND UV EMISSION FROM $1.5<\lowercase{z}<3$ STAR-FORMING GALAXIES\footnote{Based on observations made with the NASA/ESA Hubble Space Telescope, obtained from the data archive at the Space Telescope Institute. STScI is operated by the association of Universities for Research in Astronomy, Inc. under the NASA contract NAS 5-26555.}}
\author{Nicholas A. Bond, Eric Gawiser}
\affil{Physics and Astronomy Department, Rutgers University,    Piscataway, NJ 08854-8019, U.S.A.}
\email{nbond@physics.rutgers.edu}
\email{gawiser@physics.rutgers.edu}
\and
\author{Anton M. Koekemoer}
\affil{Space Telescope Science Institute, 3700 San Martin Drive, Baltimore, MD 21218, U.S.A.}
\email{koekemoer@stsci.edu}

\begin{abstract} 
We present the results of a comparative study of the rest-frame optical and rest-frame ultraviolet morphological properties of 117 star-forming galaxies (SFGs), including BX, BzK, and Lyman break galaxies with $B<24.5$, and 15 passive galaxies in the region covered by the Wide Field Camera 3 Early Release Science program.   Using the internal color dispersion (ICD) diagnostic, we find that the morphological differences between the rest-frame optical and rest-frame UV light distributions in $1.4<z<2.9$ SFGs are typically small (ICD~$ \sim 0.02$).  However, the majority are non-zero ($56$\%\ at $>3\sigma$) and larger than we find in passive galaxies at $1.4<z<2$, for which the weighted mean ICD is $0.013$.  The lack of morphological variation between individual rest-frame ultraviolet bandpasses in $z \sim 3.2$ galaxies argues against large ICDs being caused by non-uniform dust distributions.  Furthermore, the absence of a correlation between ICD and galaxy UV-optical color suggests that the non-zero ICDs in SFGs are produced by spatially distinct stellar populations with different ages.  The SFGs with the largest ICDs ($\gtrsim 0.05$) generally have complex morphologies that are both extended and asymmetric, suggesting that they are mergers-in-progress or very large galaxies in the act of formation.  We also find a correlation between half-light radius and internal color dispersion, a fact that is not reflected by the difference in half-light radii between bandpasses. In general, we find that it is better to use diagnostics like the ICD to measure the morphological properties of the difference image than it is to measure the difference in morphological properties between bandpasses.  
\end{abstract}

\keywords{cosmology: observations --- galaxies: formation -- galaxies: high-redshift -- galaxies: structure}

\vspace{0.4in}

\section{INTRODUCTION}

The Hubble Sequence of galaxies \citep{HubbleSeq} is seen out to $z \sim 1.5$ \citep{Glazebrook95,vdB96,Griffiths96,Brinchmann98,Lilly98,Simard99,vD00,Stanford04,Ravindranath04} beyond which high-redshift galaxies typically appear compact and irregular \citep[e.g.][]{Giavalisco96,Lowenthal97,Dickinson,vdB01}.   Galaxies at high redshift were initially identified using the Lyman-break technique \citep{Steidel96}, wherein $z>2.5$ galaxies are selected using a flux discontinuity in the continuum (e.g. a U-band dropout) caused by absorption from intervening neutral hydrogen \citep[e.g.][]{Rafelski09,Yan09,Oesch10,Bunker10,Hathi10}.  Subsequent techniques, such as BzK \citep{Daddi04} and `BX' \citep{Adelberger04} color selection, as well as narrow-band selection of Ly$\alpha$ emitters \citep{Hu96}, have allowed for the photometric identification of star-forming galaxies over a wider redshift range.  

The morphological properties of high-redshift galaxies are quantified using a wide range of techniques, including the concentration and asymettry indices \citep{Conselice05}, Gini coefficients \citep{Lotz06}, and S\'{e}rsic profile fitting \citep[e.g.][]{Cassata05}, among others.  At $2 \lesssim z \lesssim 3.5$, star-forming galaxy (SFG) sizes range from $< 1$~kpc to $\sim5$~kpc, with the largest often exhibiting multiple photometric components \citep[e.g.][]{Bouwens04,Ravindranath06,Oesch09}.  Morphological analyses have revealed that most of these systems are disturbed and disk-like (i.e., with exponential light profiles), with only $\sim 30\%$ having light profiles consistent with galactic spheroids \citep[e.g.,][]{Ferguson04,Lotz06,Ravindranath06,Petty09}.  Because high-redshift galaxies are typically faint and clumpy/irregular, it can be difficult to quantify their morphology in a meaningful way.  Using a sample of $97$ Lyman-$\alpha$ emitters, \citet{BondLAE} found that even the half-light radius could not be accurately measured without S/N~$\gtrsim 30$ within a $0\farcs6$ aperture.

In 2003, \citet[][P03]{Papovich03} introduced a differential morphological diagnostic known as the internal color dispersion (ICD) that could be used to quantify the morphological differences between bandpasses for astronomical objects.  When applied to galaxies, this proved to be particularly useful for comparing the rest-frame ultraviolet light distribution, which often appears clumpy even in low-redshift galaxies, to the rest-frame optical light distribution.  If the young stars in a galaxy are distributed differently from the old stars (as is usually the case at low redshift), then that galaxy will have a large ICD.  Similarly, inhomogeneous dust distributions can lead to spatial variations in the level of obscuration of rest-frame UV light, leading to a large ICD.  Their initial study focused on low-redshift galaxies and found that mid-type spirals exhibited the largest ICDs ($\xi \sim 0.2$), irregular galaxies were intermediate ($\xi \sim 0.1$), and early-type galaxies had negligible morphological variation between bands ($\xi \sim 0$).  The authors also concluded that ICDs could be reliably measured for high-redshift galaxies so long as the angular resolution ($\gtrsim 0.5$~beam~kpc$^{-1}$$\simeq0\farcs1$ at $z=2$) and signal-to-noise (S/N $\gtrsim80$) were sufficient.

Following up on this exploratory study, \citet{Papovich05}(P05) computed the ICDs of separate flux-limited galaxy samples at $z \sim 1$ and $z \sim 2.3$ using a combination of Hubble Space Telescope ({\it HST}) Wide-Field Planetary Camera 2 and Near Infrared Camera and Multi-Object Spectrometer imaging.  The authors found that the ICDs of $z \sim 1$ galaxies were generally larger and more varied than those of their higher-redshift counterparts, with the former sample including many spiral galaxies and the latter being dominated by irregulars and interacting systems.  They interpret these changes as being due to increasing stellar population heterogeneity at low redshift, with gaseous disks forming around older spheroids.  In addition, they speculated that the high-redshift galaxies with large ICDs were actually mergers-in-progress based upon their large apparent asymmetries.

With the installation of the Wide Field Camera 3 (WFC3) on {\it HST}, we now have a new, improved window into the rest-frame optical light of galaxies at $1 \lesssim z \lesssim 3$.  The deep NIR images being taken as part of the WFC3 Hubble Ultra Deep Field 2009 (GO 11563: PI Illingworth) and WFC3 Early Release Science \citep[ERS,][]{Windhorst10} programs allow for a detailed study of high-redshift galaxy morphologies down to $H_{160} \sim 26$~mag.  In this paper, we present the results of a comparative study of the rest-frame optical and rest-frame ultraviolet morphological properties of 132 SFGs in the WFC3 ERS region of GOODS-S, using the ICD as our primary diagnostic.  Throughout we will use AB magnitudes and assume a concordance cosmology with $H_0=71$~km~s$^{-1}$~Mpc$^{-1}$, $\Omega_{\rm m}=0.27$, and $\Omega_{\Lambda}=0.73$ \citep{WMAP}.  With these values, $1\arcsec = 8.2$~physical~kpc at $z=2.5$.

\section{DATA AND SAMPLE}
\label{sec:data}

Our morphological analyses are performed on star-forming and passive galaxies at $1.5<z<3.6$ using {\it HST}/ACS and {\it HST}/WFC3 imaging in the GOODS-S field.  The data products and galaxy sample are described below.

\subsection{Advanced Camera for Surveys Imaging}
\label{subsec:ACS}

In the Chandra Deep Field-South, the southern half of the GOODS survey \citep{GOODS}
covers $\sim 160$~arcmin$^2$ of sky and has {\it HST}/ACS observations in the F435W, F606W, F775W, and F850LP filters (hereafter designated $B_{435}$, $V_{606}$, $I_{775}$, and $z_{850}$).  For this study, we use only the $I_{775}$ image, which was multidrizzled to a pixel scale of $60$~mas.  The effective exposure time of this survey is variable across the GOODS area, but for $1\farcs2$ fixed aperture, a typical $I_{775}$-band, $5 \,\sigma$ detection limit is $m_{\rm AB} = 27.4$. 

\subsection{Wide Field Camera 3 Imaging}
\label{subsec:WFC3}

In order to probe the rest-frame optical light of $z\sim 2.3$ galaxies, we use {\it HST}/WFC3 F160W images (hereafter designated $H_{160}$ images) taken as part of the WFC3 ERS.  The $H_{160}$ imaging consisted of 20 orbits over ten visits to the GOODS-S field (Program 11563), leading to a total of 60 exposures.  The exposures were then reduced and calibrated (Koekemoer et al. 2010, in prep.), incorporating SPARS100 dark frames and correcting for residual gain differences between the mosaic quadrants. This also included the removal of large-scale scattered light residuals and satellite trails. The exposures were then combined using MultiDrizzle \citep{Koekemoer02} using a pixfrac of $0.8$ and a square kernel, to produce final drizzled images with a pixel scale of $60$~mas. For a $1\farcs2$ fixed aperture, a typical $H_{160}$-band, 5$\sigma$ detection limit is $m_{AB} = 27.2$.  In order to ensure accurate alignment of the IR imaging with the GOODS ACS imaging, the WFC3 exposures were individually aligned to the GOODS-S v2.0 z-band catalog and to each other. Even for single exposures, the number of sources matched to the reference catalog was $\sim 500$ and the final astrometric solution appears to be robust to $\sim 10$~mas or better. Further details on this reduction are presented in a forthcoming paper (Koekemoer et al. 2010, in prep.)

\begin{deluxetable}{lcccc}
\tablecaption{Galaxy Sample Properties\label{tab:Samples}}
\tablewidth{0pt}
\tablehead{
\colhead{Sample}
&\colhead{Number}
&\colhead{Redshift Range}
&\colhead{Reference}\\
&
&
&
}
\startdata
SFGmain	&100	&$1.4 < z < 2.9$ &1 \\
SFGhighz	&17	&$2.9 < z < 3.5$ &1 \\
PassGal	&15	&$1.4 < z < 2.0$ &2 \\
\enddata

\tablerefs{
(1) Balestra et al. 2010; (2) Cameron et al. 2010}

\end{deluxetable}

\subsection{Galaxy Sample}
\label{subsec:samples}

Our galaxy sample is broken down into three subsamples: $1.4<z<2.9$ SFGs (SFGmain), $2.9<z<3.5$ SFGs (SFGhighz), and $1.4<z<2$ passive galaxies (PassGal).  The passive galaxy subsample includes $15$ objects found in the WFC3 ERS region using a {\it YHVz} color selection technique \citep{Cameron10}.  No further cuts were made on this subsample.

The SFG subsamples are drawn from the VLT/VIMOS spectroscopic observations described in \citet{Popesso09} and \citet{Balestra10}.  The VLT/VIMOS pointings targeted high-redshift galaxies in GOODS-S with $B<24.5$, including U-band dropouts, BzK color-selected objects \citep{Daddi04}, sub-U-dropouts \citep[similar to 'BX' selection,][]{Adelberger04}, and X-ray sources in the \citet{Giacconi02} and \citet{Lehmer05} catalogs.  For this analysis, we exclude objects with redshift quality flag C (poor quality) and objects satisfying the BzK color selection for passively-evolving galaxies.  In order to exclude unobscured AGN from our subsample of SFGs, we remove any object targeted as an X-ray source by \citet{Popesso09}, as well as any object that is within $2$~\arcsec\ of an X-ray source in the expanded catalogs of \citet{Virani06} and \citet{Luo08}.

Of the remaining SFGs in the WFC3 ERS region, $100$ lie within $1.4<z<2.9$ and are assigned to SFGmain, while a further $17$ objects are within $2.9<z<3.6$ and are assigned to SFGhighz.  This separation is motivated by P03, who found that the internal color dispersion is most effective at identifying the signatures of heterogeneous stellar populations when the two filters cover opposite sides of the Balmer and $4000$~\AA\ breaks.  Since we are using images taken with the $H_{160}$ and $I_{775}$ filters on HST, the light from young stars will dominate both images for galaxies at $z\gtrsim2.9$ (that is, those in SFGhighz) and we only expect to see a large ICD in one of these galaxies if there is a large column of inhomogeneously distributed dust.  

The galaxy subsamples are described in Table~\ref{tab:Samples}.  For more detail on the selection of these subsamples, see the references given in the table.   The sky positions and redshifts of individual objects in our parent sample are listed in Table~\ref{tab:MainObj}, ordered by catalog number.  In this paper, we will refer to SFGs according to their line number in {\it version 2.0} of the VIMOS master catalog \citep{Balestra10}, excepting the passive galaxies, which will be denoted P$1$-P$15$.  Although it is listed in Table~\ref{tab:MainObj}, we exclude object 1379 from the majority of our analysis because it was not detected in either the $H_{160}$ image or the $I_{775}$ image.

\begin{deluxetable}{lcccc}
\tablecaption{Galaxy Samples\label{tab:MainObj}}
\tablewidth{0pt}
\tablehead{
\colhead{Number}
&\colhead{Sample}
&\colhead{$\alpha$}
&\colhead{$\delta$}
&\colhead{$z$}
\\
&
&
&
&
}
\startdata
$447$ &SFGmain &$3$:$32$:$01.117$ &$-27$:$44$:$04.830$ &$2.6355$  \\
$455$ &SFGmain &$3$:$32$:$01.302$ &$-27$:$42$:$43.973$ &$2.2998$  \\
$462$ &SFGmain &$3$:$32$:$01.509$ &$-27$:$44$:$22.574$ &$2.5119$  \\
$479$ &SFGmain &$3$:$32$:$02.067$ &$-27$:$45$:$13.878$ &$2.3131$  \\
$482$ &SFGhighz &$3$:$32$:$02.167$ &$-27$:$42$:$30.428$ &$3.2630$  \\
\enddata
\tablenotetext{*}{This table is only a stub.  A manuscript with complete tables is available at http://www.nicholasbond.com/Bond1004.pdf}

\end{deluxetable}

\section{METHODOLOGY}
\label{sec:method}

In the following section, we describle our methodology for the measurement of fixed-aperture quantities, including half-light radius and light centroid, as well as the internal color dispersion, a differential diagnostic of the morphological differences between two bandpasses.  Our analysis is based upon techniques presented in P03, P05, and \citet{BondLAE}. 

\subsection{Fixed-aperture magnitudes and half-light radii}
\label{subsec:magrads}

Because star-forming galaxies are often clumpy, consist of multiple components, and may be interacting with nearby galaxies, there is some ambiguity in choosing the appropriate aperture for morphological analysis.  We will use a simple approach in this paper, measuring all quantities (with the exception of the internal color dispersion, see Section~\ref{subsec:ICD}) within a fixed $1\farcs2$ aperture.  

Our methodology for measuring object magnitudes, centroids, and fixed-aperture half-light radii is described in detail in \citet{BondLAE}.  All of these measurements are performed on the original drizzled images, prior to the point spread function (PSF) convolution described below.  

Using Source Extraction software \citep[SExtractor,][]{SExtractor}, we estimate the maximum contamination by interloping sources within the $1\farcs2$ apertures to be $\sim 17$\% based on the background sky density of sources with $H_{160}<26$.  However, because the SFGs in our sample were selected using ground-based photometry, the contamination rate is probably much lower because an interloper within $\sim 1$\arcsec\ of a high-redshift galaxy would have blended with it and changed its apparent color.  Since these ``contaminated'' high-redshift galaxies will sometimes fall outside of the high-redshift galaxy color selection regions, they are less likely to be selected than their uncontaminated counterparts.

\subsection{Internal color dispersion}
\label{subsec:ICD}

\begin{figure}
\epsscale{1.3}
\plotone{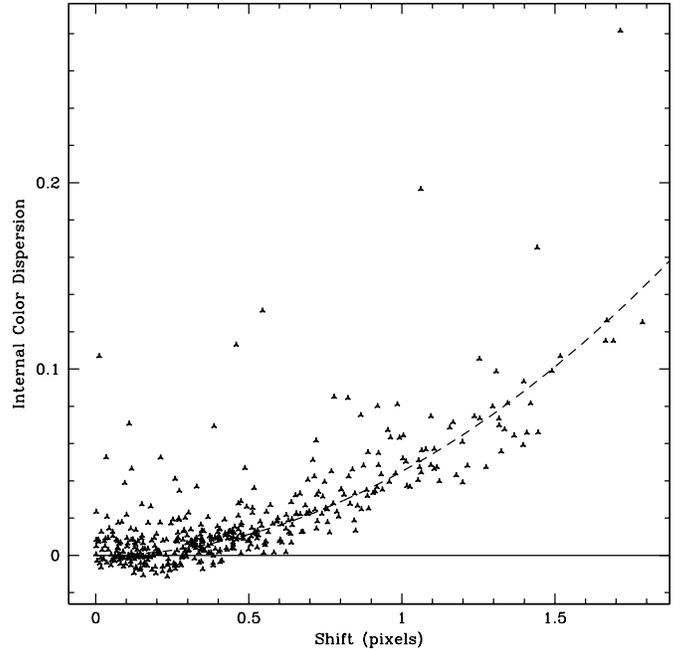}
\caption{Internal color dispersion as a function of centroid offset between $I_{775}$ and $H_{160}$-band images  for a series of Monte Carlo simulations of an SFG in the WFC3 ERS, where each point is computed with a random realization of the sky noise in the images.  The dashed line indicates an iterative quadratic fit to the data (discarding outliers).
\label{fig:ICD_vs_Offset}}
\end{figure}

The internal color dispersion attempts to quantify the differences in the distribution of an object's light between two observed bandpasses without making model-dependent assumptions.   When comparing between the rest-UV and rest-optical, large ICDs can arise if the young stars in a galaxy are distributed differently from the old stars or the dust is inhomogeneously distributed within the galaxy, leading to spatial variations in the level of obscuration of rest-frame UV light.  At low redshift, P03 found that irregular galaxies have $\xi\sim0.1$, while mid-type spiral galaxies have $\xi\sim0.2$.  High-redshift star-forming galaxies are morphologically similar to irregular galaxies, but have had less time to form previous generations of stars, so the majority would be expected to have $\xi<0.1$.

For a noise-free image, the ICD is given by,
\begin{equation}
\xi(I_1,I_2)=\frac{\sum\limits_{i=0}^{N}(I_{2,i}-\alpha I_{1,i}-\beta)^2}{\sum\limits_{i=0}^{N}(I_{2,i}-\beta)^2},
\label{eq:ICDtheory}
\end{equation}
where the sum is performed over $N$ pixels in a chosen aperture, $I_{1,i}$ and $I_{2,i}$ are the fluxes in pixel $i$, $\alpha$ is the ratio of the total fluxes between cutout 2 and cutout 1, and $\beta$ is the difference in the background between the two cutouts.  Since we perform SExtractor sky subtraction on all cutouts prior to analysis \citep[see][]{BondLAE}, we will assume $\beta=0$ for all of our measurements.

If we wish to accurately compute a differential morphological diagnostic between two observed bandpasses, we must ensure that (1) the effective PSF is the same in the two images and (2) there is no astrometric offset.  If the PSF is well approximated by a Gaussian in both frames, the former requirement can be satisfied by convolving the lower-resolution image with a Gaussian that has $\sigma=\sqrt{\sigma_2^2-\sigma_1^2}$.  Unfortunately, the PSF of the WFC3 $H_{160}$-band imaging is not well approximated by a Gaussian \citep{WFC3Handbook}, so we must convolve each image with the PSF of the other.  We estimated the PSF of each image using a stack of stars from \citet{MartinStars} that fell within the WFC3 ERS region and had temperatures consistent with $K$ and $M$-type stars, a regime in which photometric confusion with galaxies is minimal.  We further restricted our stack to $29$ stars that were isolated (i.e., the only photometric component within $0\farcs6$), unambiguously stellar (SExtractor stellarity $>0.9$), and unsaturated.  Experiments with stars of a variety of spectral types reveal that the internal color dispersion is insensitive to the spectral shape of the PSF across the $H_{160}$ filter, with ICDs varying by $<10$\%\ of the typical ICD uncertainty in SFGmain.  

Noise contributions from the sky background in images of high redshift galaxies will affect the terms in Equation~\ref{eq:ICDtheory}.  In order to estimate the amplitude of this contribution, we extracted $400$  $5$\arcsec$\times 5$\arcsec\ cutouts from random positions on the $H_{160}$ and $I_{775}$ images.  Cutouts were discarded if they contained any detected sources.  We then computed a noise term,
\begin{align}
\xi_n(\alpha,N)=\mathrm{Median}(\sum\limits_{i=0}^{N}I_{2,i}^2)-2\alpha\,\mathrm{Median}(\sum\limits_{i=0}^{N}I_{1,i}I_{2,i}) \nonumber \\
+\alpha^2\,\mathrm{Median}(\sum\limits_{i=0}^{N}I_{1,i}^2),
\label{eq:ICDnoise}
\end{align}
where the medians were computed over the set of blank cutouts.  This was then subtracted from the numerator and denominator of Equation~\ref{eq:ICDtheory} when computing the ICDs of the objects in our sample.

We calculated the ICDs using $0\farcs6$ to $1\farcs5$ circular apertures on $5$\arcsec$\times 5$\arcsec\ cutouts in the {\it HST}/WFC3 and {\it HST}/ACS images, each centered at the flux-weighted centroid of all SExtractor detections within $1\farcs2$ of the object positions reported in \citet{Balestra10} and \citet{Cameron10}.  The sizes of the apertures used to compute the ICDs for individual galaxies were selected on an object-by-object basis based upon a visual inspection of the cutouts.

\subsection{Monte Carlo Simulations}
\label{subsec:Simulations}

\begin{figure}
\epsscale{1.3}
\plotone{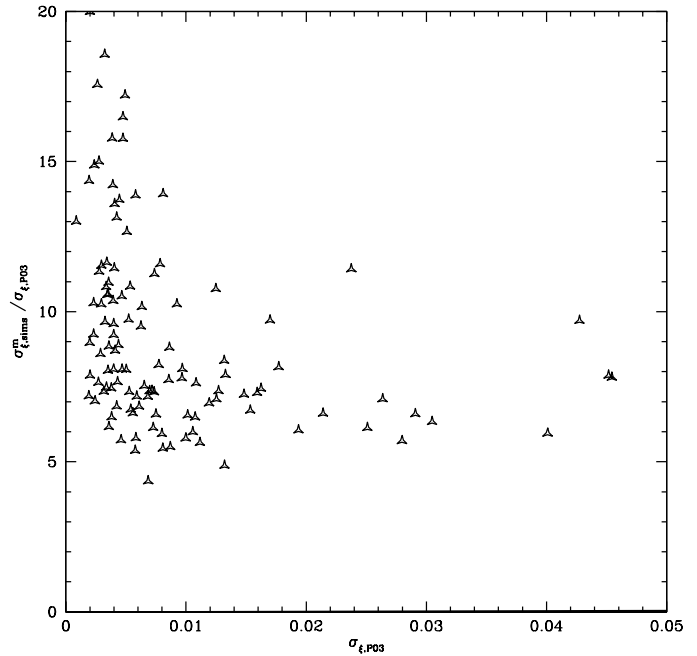}
\caption{Ratio between the uncertainty in internal color dispersion in our Monte Carlo simulations and the uncertainty as determined by the analytical approximation of P03 (see Equation~\ref{eq:deltaP03}) for the objects in our $z \sim 2.5$ SFG subsample.  All points have $\sigma^m_{\xi,sims}/\sigma_{\xi,P05}>1$ as a result of correlated noise in the drizzled HST images, but the analytical approximation becomes particularly bad at high signal-to-noise (i.e. small $\sigma_{\xi}$).
\label{fig:ErrorFactor}}
\end{figure}

For this study, we estimate the uncertainties on the internal color dispersion measurements for individual galaxies in our sample using object-by-object Monte Carlo simulations.  In order to do this, we first extracted the galaxies in our sample from their $H_{160}$ cutout using SExtractor (DETECT\_MINAREA~$=5$ and DETECT\_THRESH~$=1.65$).  Mock $H_{160}$ cutouts were then created by adding the extracted light profile to cutouts centered at random positions on the $H_{160}$ image (see Section~\ref{subsec:ICD}).  We used the same $H_{160}$-extracted galaxy light distributions to generate mock $I_{775}$ cutouts, normalizing the light distribution such that the detected flux was equal to that in the real $I_{775}$ image.  This accounts for the object's $I_{775}-H_{160}$ color.   Finally, we added noise realizations from the $I_{775}$ image.

The other significant source of uncertainty in our measurements is the unknown astrometric shift between the WFC3 and ACS imaging.  In order to simulate this, we applied random two-dimensional shifts to the mock $H_{160}$ cutouts, with a mean amplitude of $10$~mas (our astrometric uncertainties, see Section~\ref{subsec:WFC3}).

Our simulations are designed to approximate the scatter induced by the sky background on ICD measurements of an object with an intrinsic $\xi=0$ and will allow us to distinguish objects with intrinsic non-zero ICDs.  Systematic sources of error, such as astrometric offsets and non-uniform sky backgrounds, only produce positive scatter, so the error distribution is non-Gaussian and we estimated the bias and positive scatter using the median and third quartile, where we define,
\begin{equation}
\sigma_\xi^m=1.496[{\mathrm Q3}(\xi)-{\mathrm{Median}}(\xi)].
\label{eq:sigmaxi}
\end{equation}
We performed a total of $400$ simulations for each object.

Figure~\ref{fig:ICD_vs_Offset} shows the dependence of the ICD on the relative astrometric shift between $I_{775}$ and $H_{160}$ for one of our simulations.  The dashed line indicates an iterative quadratic fit to the data (discarding outliers).  Each measurement includes random realizations of the sky noise, leading to the scatter seen about the best-fit curve.  For the objects in our sample, the median astrometric shift at which the resulting systematic error in the ICD exceeds the random ICD uncertainty is $\sim 0.64$~pixels, although the precise behavior depends upon the shape of the light distribution.  Point sources, for example, will exhibit more sensitivity to astrometric errors than well-resolved galaxies.

In Figure~\ref{fig:ErrorFactor}, we compare the simulated uncertainties for the SFGs in our sample to those that would be derived from the analytical approximation of P03,
\begin{equation}
\delta(\xi)=\frac{\sqrt{2/N_{{\rm pix}}}\sum\limits_{i=0}^{N}(B_{2,i}^2+\alpha^2B_{1,i}^2)}{\sum\limits_{i=0}^{N}(I_{2,i}-\beta)^2-\sum\limits_{i=0}^{N}(B_{2,i}-\alpha B_{1,i})^2},
\label{eq:deltaP03}
\end{equation}
where $B_{1,i}$ and $B_{2,i}$ are the fluxes in pixel $i$ in a cutout extracted from a blank region of sky.  As the expression shows, large ICD uncertainties are generally found when a source is detected at low signal-to-noise in one of the images.  At high redshift, the morphological variations that we see between wavebands are small, even for galaxies with a lot of dynamical activity, so sources with individual detections that have S/N $\gtrsim 30$ can still have uncertainties on their ICD comparable to or larger than the median ICD in our sample ($\xi \sim 0.02$).  As the authors of P03 note, their expression will underestimate the uncertainties in the presence of correlated noise.  Since {\it HST} images are typically drizzled to correct for oversampling of the PSF, our simulations do in fact yield larger uncertainties than predicted by P03.  P05 accounted for this discrepancy in drizzled images by correcting the uncertainty by a single empirically-determined multiplicative factor, $C(\xi)$.  Although $C(\xi)=6$ would be a reasonable approximation at $\sigma_\xi \gtrsim 0.01$, it would underestimate the uncertainty for some of the brighter objects in our sample ($\sigma_{\xi,P05}\lesssim0.01$) by factors up to $\sim 3$.  Hereafter, we will only use the uncertainties derived from our Monte Carlo simulations.

\begin{figure}
\epsscale{1.2}
\figurenum{3}
\plotone{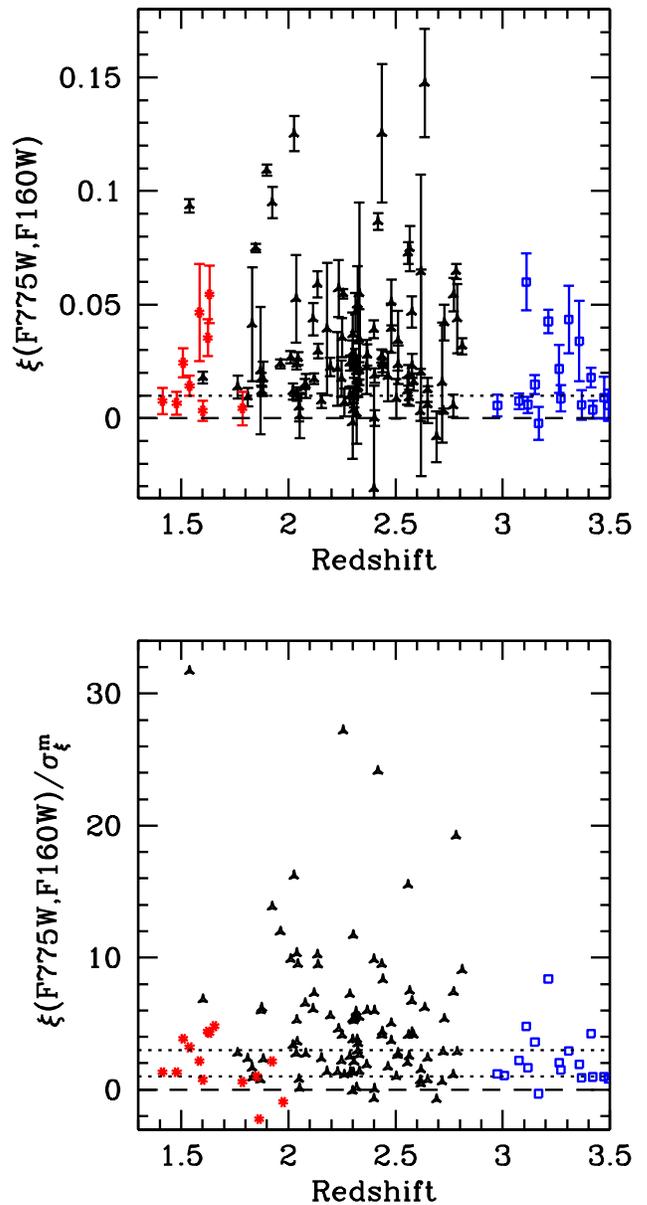}
\caption{Internal color dispersion (top) and ICD signal-to-noise (bottom) as a function of redshift.  We indicate $1.4<z<2.9$ SFGs with triangles, $2.9<z<3.5$ SFGs with open squares, and $1.4 < z < 2$ passive galaxies with stars.  All error bars are drawn from our Monte Carlo simulations and galaxies with $\sigma_\xi^m>0.05$ are not plotted.  In the top panel, the dotted line indicates $\xi=0.01$ and in the bottom panel, the two dotted lines indicate an ICD signal-to-noise of 1 and 3.  Objects with $\xi(F775W,F160W)/\sigma^m_{\xi}>3$ show statistically significant morphological differences between the $I_{775}$ and $H_{160}$-band images.  While the majority of star-forming galaxies at $1.4<z<2.9$ lie above this threshold, very few passive galaxies or $z>3$ galaxies have $\xi>0$ at the depth and resolution of these images.
\label{fig:Disp_vs_z}}
\end{figure}

\section{RESULTS}
\label{sec:results}

Below we present the results of a morphological analysis of the high-redshift star-forming and passive galaxies in our sample.  All measured quantities, including centroids, magnitudes, half-light radii, and ICDs, of these galaxies are compiled in Table~\ref{tab:ObjPhot}.

\subsection{Internal color dispersion}
\label{subsec:ICDResults}

\subsubsection{Star-forming galaxies at $1.4<z<2.9$ }
\label{subsubsec:ICDSFGmain}

Some example $I_{775}$ and $H_{160}$ cutouts for $15$ SFGs with $\xi>0.05$ are shown in Figure~\ref{fig:Cutouts_LargeICD}.  Of the galaxies shown, all are in SFGmain except for object 863, which was identified as a Ly$\alpha$ emitter (LAE) at $z=3.1$ in \citet{GronwallLAE}.  While it is possible that the extended V-band emission seen in the cutout originates from the same high-redshift source as the Ly$\alpha$ emission, the large ICD, distinct disk-like morphology, and large size compared to the majority of LAEs at that redshift \citep{BondLAE} suggest that it may be a low-redshift galaxy superimposed on a high-redshift LAE.  The remaining galaxies exhibit complex morphologies, many with multiple components in both the rest-frame optical and rest-frame ultraviolet.  Some of the objects, such as 1204 and 1598, resemble the ``chain galaxies'' of \citet{CHS95}, which are believed to be massive galaxies in the act of formation, while others (447 and 560) have two-component morphologies suggestive of mergers.

The distribution of internal color dispersions is shown as a function of redshift in Figure~\ref{fig:Disp_vs_z}.  The $100$ galaxies in SFGmain exhibit a wide range of ICDs, $56$ of which are non-zero at $>3\sigma$ (see bottom panel).  Of the galaxies with small uncertainties in the ICD ($\sigma_\xi^m<0.01$), only $22$\%\ have $\xi<0.01$, suggesting that the majority of color-selected SFGs in this redshift range have underlying populations of old stars that are spatially distinct from the young stars.  The median ICD in this subsample is $0.023$.

There is no apparent correlation between $\xi$ and $z$ in SFGmain, except for a possible downturn at $z \gtrsim 2.6$.  It may be that the ICD becomes less sensitive to stellar population heterogeneity as the $H_{160}$ filter approaches the rest-frame $4000$~\AA\ break.  Alternatively, it may be due to evolution - galaxies are expected to evolve towards less stellar population heterogeneity and lower dust mass with increasing redshift \citep{Hayes10}.  There is a small subset of 21 SFGs with a very large ICD ($\xi>0.05$), which appear to make up a much larger fraction of the SFG subsample at $z<2$.  This is consistent with the findings of P05, who found that galaxies with very large ICDs went from $7$\%\ of their $H$-band-selected sample at $z \sim 2.3$ to $25$\%\ at $z \sim 1$.

\begin{figure}
\epsscale{1.3}
\figurenum{4}
\plotone{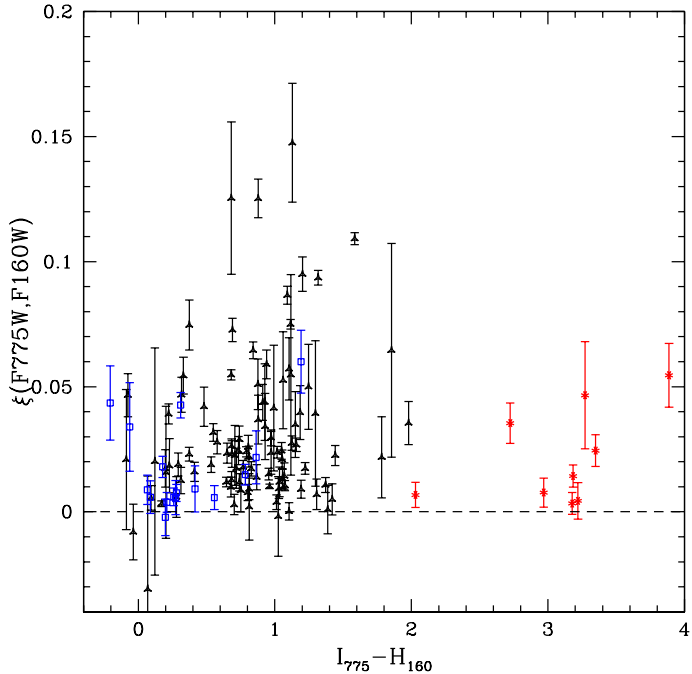}
\caption{Internal color dispersion as a function of $I_{775}-H_{160}$ color.  Points and error bars are the same as in Figure~\ref{fig:Disp_vs_z} and galaxies with $\sigma_\xi^m>0.05$ are not plotted.  There is no evidence of a correlation between ICD and rest-UV-to-rest-optical color in the $2.9<z<3.5$ SFGs (black points).
\label{fig:Disp_vs_Color}}
\end{figure}

Figure~\ref{fig:Disp_vs_Color} plots $\xi$ as a function of $I_{775}-H_{160}$ color.  We find no evidence for a correlation in SFGmain.  Although this is in agreement with the bulk properties of the P05 sample, they found that the galaxies with very large ICDs had very red $I_{775}-H_{160}$ colors.  We see no evidence for this effect in SFGmain, but considering that P05 only had 7 galaxies with $\xi>0.05$, the apparent differences may be due solely to small-number statistics.  The lack of a correlation between $\xi$ and $I_{775}-H_{160}$ supports the hypothesis that the morphological differences between the rest-frame ultraviolet and rest-frame optical are due primarily to stellar population heterogeneity rather than inhomogeneous dust distributions, as greater dust columns would both redden the galaxies and produce larger ICDs.  A large population of old stars can also redden a galaxy's $I_{775}-H_{160}$ color, but a galaxy with a larger fraction of old stars will only have a larger ICD if the two populations are spatially segregated and the old stars are not dominating the emission in the $I_{775}$ and $H_{160}$ bands.

\subsubsection{Passive galaxies}
\label{subsubsec:ICDSFGPassive}

\begin{figure}
\epsscale{1.2}
\figurenum{9}
\plotone{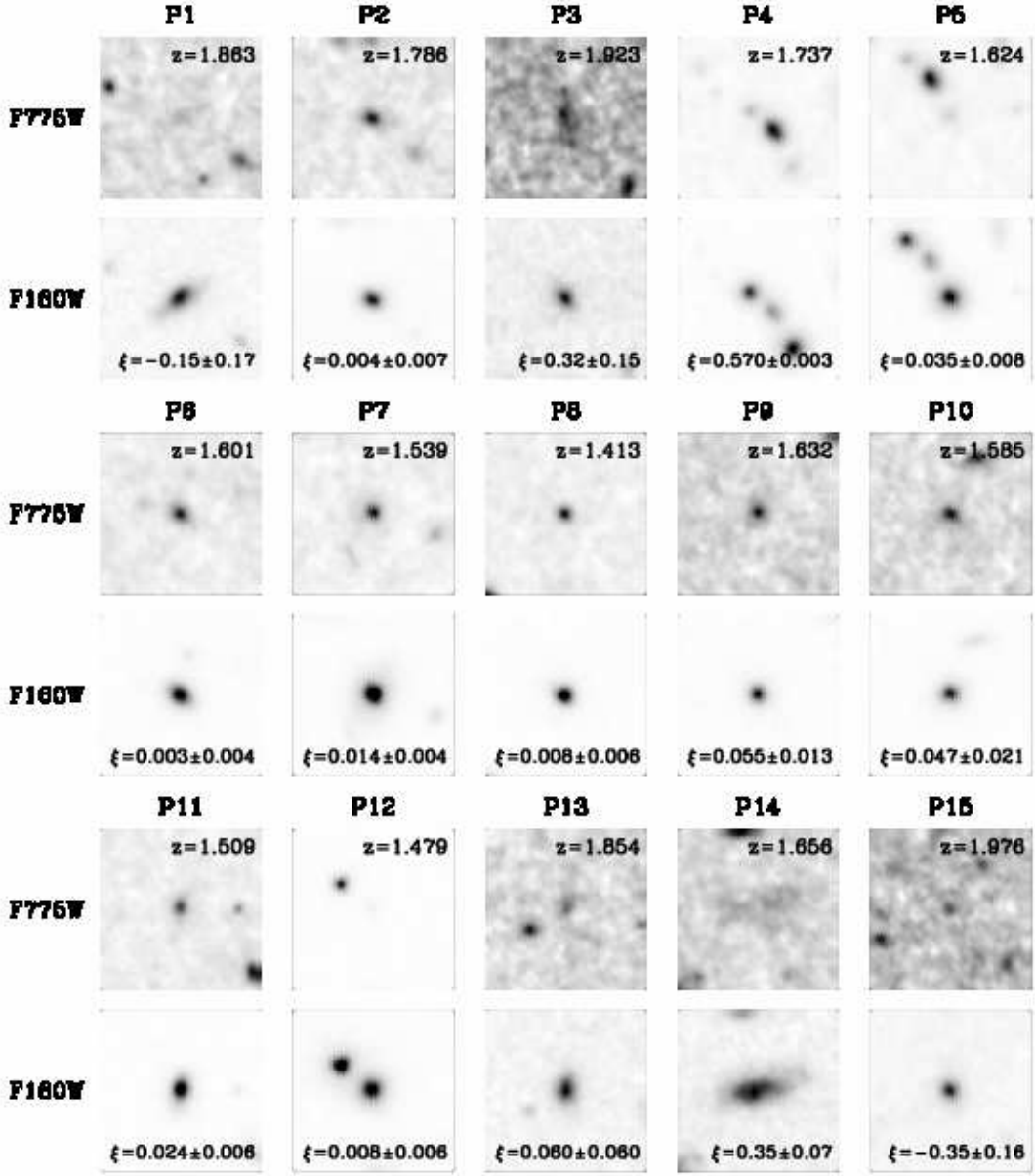}
\caption{$H_{160}$ and $I_{775}$ cutouts of 15 passive galaxies.  The cutouts are $1\farcs8$ ($30$~pixels) on a side and have been convolved with the PSF of the comparison image (i.e. $H_{160}$ with $I_{775}$ and vice versa).
\label{fig:Cutouts_Passive}}
\end{figure}

The $I_{775}$ and $H_{160}$ cutouts for the $15$ passive galaxies in our PassGal subsample are shown in Figure~\ref{fig:Cutouts_Passive}.  Objects P$1$, P$3$, and P$15$ are faint in $I_{775}$ and have $\sigma_{\xi}^m>0.1$, so there is little we can say about their differential morphologies.  There may be a diffuse component to the rest-frame ultraviolet light of P$3$ that is not present in the rest-frame optical, but the ICD is only greater than zero at the $\sim 2\sigma$ ($\xi \simeq 0.3$) level.  The $I_{775}$ light distribution of P$14$ is also faint ($\sigma_{\xi}^m = 0.07$), but the ICD is greater than zero at $\sim 5\sigma$.  Its rest-frame optical light distribution appears disk-like with a compact central component that does not appear in the rest-frame ultraviolet image, possibly suggesting the presence of a large column of dust in the galaxy center.  Object P$4$ is bright in both bandpasses, but has two components within $1\farcs2$ with very different $I_{775}-H_{160}$ colors, yielding an anomalously large ICD.  The top left component has colors consistent with a passive galaxy \citep{Cameron10} and the bottom right component may be a low-redshift interloper or a nearby star-forming galaxy.

The remaining $10$ objects have single-component, compact morphologies that appear similar in the two bandpasses.  Of the objects with $\sigma_\xi^m<0.01$, $4$ of $8$ also have $\xi<0.01$.  For comparison, only $16$ of $74$ objects in SFGmain with $\sigma_\xi^m<0.01$ have $\xi<0.01$, suggesting that the stellar populations in the passively-selected galaxies are more uniformly distributed than those in SFGs.

\subsubsection{Star-forming galaxies at $2.9<z<3.5$ }
\label{subsubsec:ICDSFGhighz}

For galaxies with $2.9<z<3.5$, the $I_{775}$ and $H_{160}$ bands both cover light blueward of the $4000$~\AA\ break and the ICD becomes a poor tracer of stellar population heterogeneity.  If the ICDs of the SFGmain subsample are dominated by this heterogeneity (as opposed to dust inhomogeneity), we expect to see a sharp dropoff in the ICDs of the SFGhighz subsample.  This dropoff does appear in the top panel of Figure~\ref{fig:Disp_vs_z}, where 8 of 11 galaxies with $\sigma_\xi^m<0.01$ also have $\xi<0.01$.  Of the entire subsample of 16 galaxies, only 4 have $\xi>0$ at $>3\sigma$ and only 1 at $>5\sigma$.

\subsection{Half-light radii and centroid offsets}
\label{subsec:hlr}

\begin{figure}
\epsscale{1.3}
\figurenum{5}
\plotone{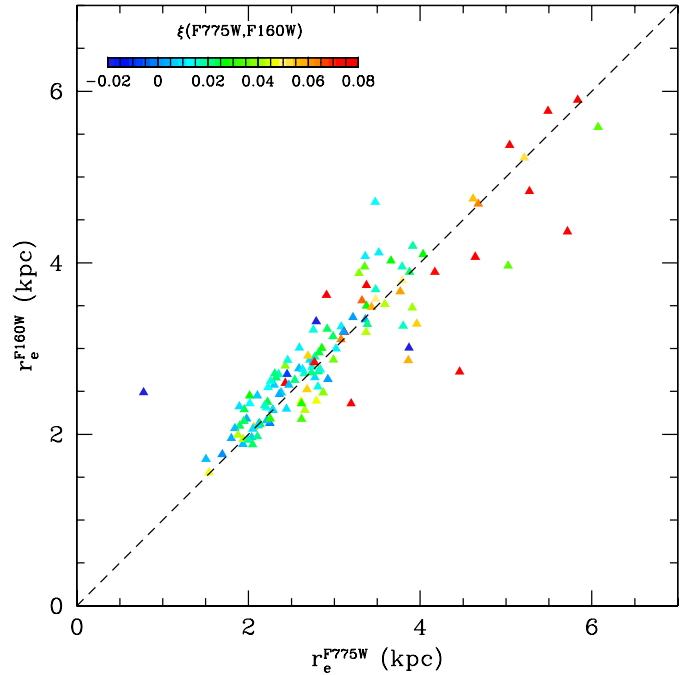}
\caption{Half-light radius in $H_{160}$ as a function of half-light radius in $I_{775}$ for SFGs at $1.4<z<2.9$, where points are color-coded according to their internal color dispersion.  The half-light radii are similar between the two bandpasses ($\sim 20$\% scatter), with a median ratio of $r_e^{F160W}/r_e^{F775W}=1.02$.  Objects sometimes show strong morphological variation between bandpasses, as indicated by a large ICD, but this is not generically accompanied by a large difference in half-light radius.
\label{fig:RvsR}}
\end{figure}

\begin{figure}
\epsscale{1.3}
\figurenum{6}
\plotone{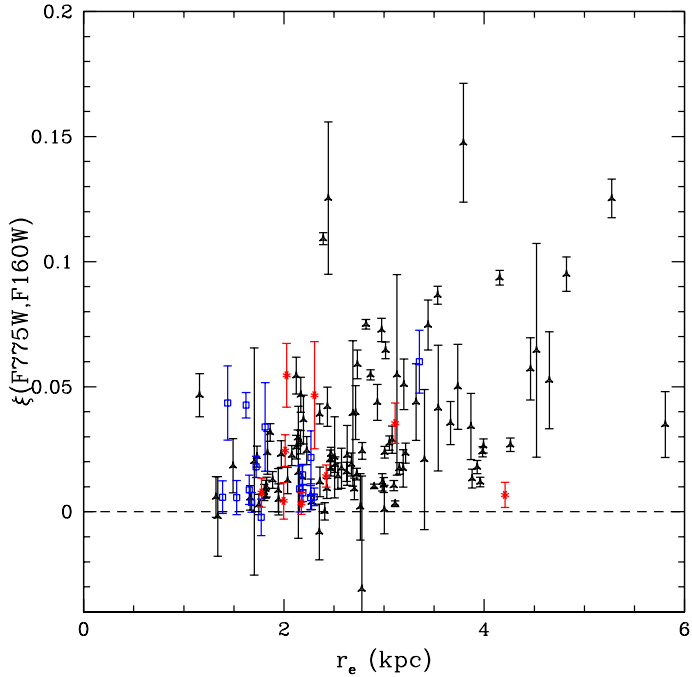}
\caption{Internal color dispersion as a function of fixed-aperture half-light radius in $H_{160}$.  Points and error bars are the same as in Figure~\ref{fig:Disp_vs_z} and galaxies with $\sigma_\xi^m>0.05$ are not plotted.  For $2.9<z<3.5$ SFGs, there is a clear increase in the mean ICD as $r_e$ increases, even for objects much larger than the WFC3 resolution limit ($r_e\sim1$~kpc), possibly indicating that large ICDs are arising from systems that are not yet dynamically settled.
\label{fig:ICD_vs_R}}
\end{figure}

High-redshift star-forming galaxies sometimes show strong morphological variation between bandpasses, as indicated by a large ICD, but differential diagnostics like the ICD are only useful if they provide us with information that single-band diagnostics, such as half-light radius and centroid offset, cannot provide.  In Figure~\ref{fig:RvsR}, we plot the rest-frame optical half-light radius as a function of the rest-frame UV half-light radius for the SFGs in our sample.  In most cases, the half-light radii differ by $\lesssim 20$\%, with the rest-frame UV light typically being more extended for the larger objects.  The typical SFG is marginally larger ($\sim 5$\%) in the $H_{160}$ image even after PSF convolution.  The points in Figure~\ref{fig:RvsR} are also color coded according to the magnitude of the ICD for that object (blue is smallest, red is largest).  Although the objects with the largest difference in half-light radius do generally have large ICDs, there are many objects that exhibit a great deal of morphological variation between bandpasses with relatively little difference in half-light radius.  Figure~\ref{fig:ICD_vs_R} indicates a clear increase in the median ICD with increasing rest-frame optical half-light radius.  Although convolution with the PSF will artificially decrease the ICD for objects near the resolution limit ($r_e\lesssim2$~kpc), the trend continues to larger radii.  The largest objects ($>3$~kpc) have a median ICD and dispersion that is $\sim 2.5$ times that of SFGs with $2<r_e<3$~kpc.  The correlation between ICD and half-light radius may indicate that the non-zero ICDs are being caused by star formation from gas that is
dynamically young; that is, gas that is spatially segregated from a dynamically older
stellar population and has been recently accreted onto
the galaxy's host dark matter halo.  By contrast, clumpy star formation that
arises from, for example, disk instabilities or stellar feedback,
would not act to increase a galaxy's size in the rest-frame optical.

\begin{figure}
\epsscale{1.3}
\figurenum{7}
\plotone{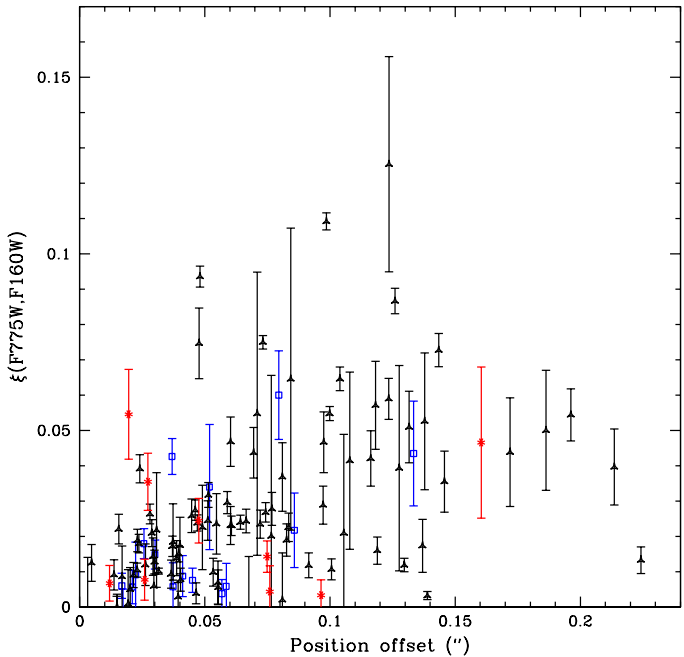}
\caption{Internal color dispersion as a function of centroid offset between the $H_{160}$ and $I_{775}$ images for a sample of galaxies in the WFC3 ERS.  Points and error bars are the same as in Figure~\ref{fig:Disp_vs_z} and galaxies with $\sigma_\xi^m>0.05$ are not plotted.  Although objects are more likely to have a large position offset when their ICD is large, the two quantities are not tightly correlated and there are many objects with $\xi>0.01$ and an offset $<0\farcs05$.
\label{fig:ICD_vs_Offset_SFG}}
\end{figure}

Unlike half-light radius, the centroid offset does appear to give some indication of the morphological variation between bandpasses (see Figure~\ref{fig:ICD_vs_Offset_SFG}).  However, there is a great deal of scatter, even out to the largest centroid offsets.  Galaxies with the largest ICDs ($\xi>0.05$) typically have centroid offsets of $\sim 0\farcs1$, corresponding to $0.83$~kpc at $z = 2.3$.

\section{DISCUSSION}
\label{sec:discussion}

The morphological differences between the rest-frame optical and rest-frame UV light distributions in $1.4<z<2.9$ SFGs are typically small ($\xi \sim 0.02$), but are statistically significant (more than half are non-zero at $>3\sigma$) and are larger than we find in the majority of passive galaxies at $1.4<z<2$.  We find that these differences are due to heterogeneous stellar populations, in agreement with P05, both because the ICDs show no correlation with UV-optical color and because a sample of higher-redshift galaxies shows much smaller ICDs (typically $\xi<0.01$) when compared between two rest-frame UV bandpasses.  We cannot rule out a contribution to the ICD from a non-uniform dust distribution, but these two facts suggest that it is sub-dominant.

\begin{figure}
\epsscale{1.2}
\figurenum{8}
\plotone{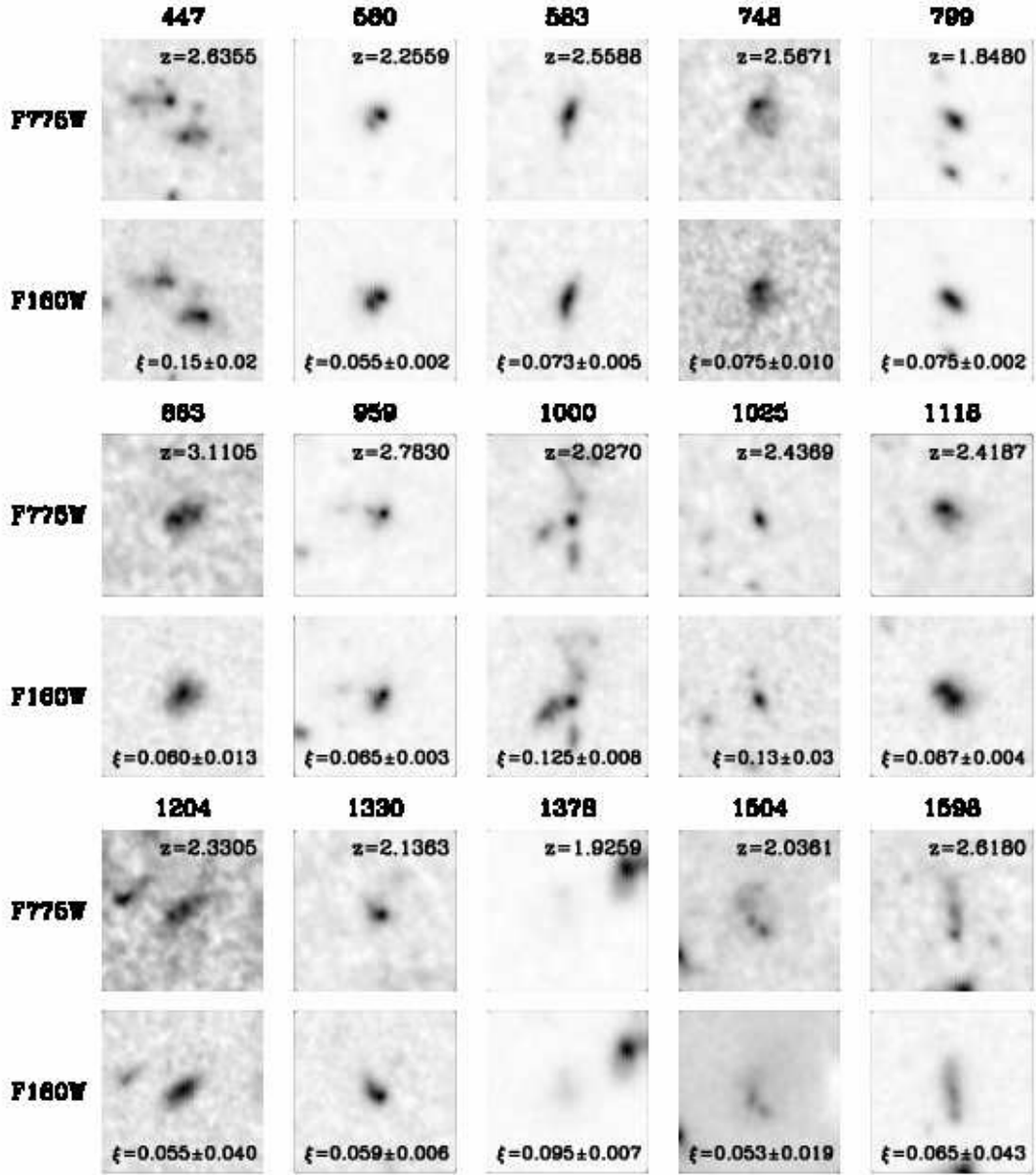}
\caption{$H_{160}$ and $I_{775}$ cutouts of 15 $1.4<z<2.9$ star-forming galaxies with $\xi>0.05$.  The cutouts are $1\farcs8$ ($30$~pixels) on a side and have been convolved with the PSF of the comparison image (i.e. $H_{160}$ with $I_{775}$ and vice versa).
\label{fig:Cutouts_LargeICD}}
\end{figure}

The SFGs with the largest internal color dispersion ($\gtrsim 0.05$, see Figure~\ref{fig:Cutouts_LargeICD}) generally have complex morphologies that are both extended and asymmetric, suggesting that they may be mergers-in-progress.  This hypothesis is difficult to prove with imaging alone, but the fact that many of these galaxies appear clumpy in the rest-frame optical suggests that it is not just the young stars that have inhomogeneous distributions, but also older generations of stars that would have had time to mix in a single, steadily-accreting system.  Furthermore, P05 showed that single galaxies with half-light radii $>3$~kpc are rare at $z \sim 2$, and that $\sim 50$\%\ of the objects in their sample with $\xi>0.08$ are larger than $3$~kpc.  If the large ICDs are simply due to clumpy star formation, it is occurring on a scale much larger than is typical for galaxies at this redshift.

Although many of the passive galaxies in our PassGal subsample are centrally compact and have ICDs that are consistent with zero, in line with the conventional picture of this class of galaxies \citep{vD04,Daddi05,Cimatti08,Franx08,Damjanov09,Szomoru10}, at least five of the galaxies have non-zero ICDs at $>3\sigma$.  This suggests that a significant fraction of largely passive galaxies at this epoch are still actively forming stars, possibly in clumpy stellar disks \citep{Stockton08,McGrath08,Cassata10}.

It is important that we continue to explore the wavelength dependence of high-redshift galaxy morphologies, as different regions of the spectrum probe different physical components of the host galaxy.  Because high-redshift galaxies are often clumpy and irregular, diagnostics such as the half-light radius and concentration index are good for obtaining a quantitative measure of the morphology in a single band but are usually poor at capturing the filter-to-filter variations (e.g., see Figure~\ref{fig:RvsR}).  As a general rule for galaxies at $z\gtrsim2$, it is better to measure the morphological properties of the difference image than the difference in morphological properties between bandpasses.  The internal color dispersion, which gives the root mean squared amplitude of the difference image, is one example of this, but any morphological diagnostic that can be applied to a single-band light distribution can also be applied to a two-band difference image.  An obvious next step would be to measure higher-order moments of the differential light distribution, but given the relatively small values of $\xi$ we find in high-redshift galaxies, one would likely require greater imaging depth in both bands to draw meaningful conclusions.

\acknowledgments

We thank Casey Papovich and Swara Ravindranath for their helpful comments on the implementation of the internal color dispersion and Martin Altmann for the use of his sample of stars in the ECDF-S.  Support for this work was provided by NASA through grant number HST-AR-11253.01-A from the Space Telescope Science Institute, which is operated by AURA, Inc., under NASA contract NAS 5-26555 and by the National Science Foundation under grant AST-0807570.

\bibliographystyle{apj}                       

\bibliography{apj-jour,Bond0106}

\begin{deluxetable}{lcccccccccc}
\tablecaption{Photometric Properties\label{tab:ObjPhot}}
\tablewidth{0pt}
\tablehead{
\colhead{Number}
&\colhead{Sample}
&\colhead{$\alpha$\tablenotemark{a}}
&\colhead{$\delta$\tablenotemark{a}}
&\colhead{$z$}
&\colhead{$H_{160}$}
&\colhead{$I_{775}$}
&\colhead{$r_{H}$\tablenotemark{b}}
&\colhead{$r_{I}$\tablenotemark{b}}
&\colhead{$\xi$\tablenotemark{c}}
&\colhead{$\sigma_\xi^m$\tablenotemark{~d}}
\\
&
&
&
&
&\colhead{(AB mags)}
&\colhead{(AB mags)}
&\colhead{(\arcsec)}
&\colhead{(\arcsec)}
&
&
}
\startdata
$447$ &SFGmain &$3$:$32$:$01.102$ &$-27$:$44$:$04.869$ &$2.635$ &$23.225 \pm 0.026$ &$24.354 \pm 0.051$ &$0.46$ &$0.43$ &$0.148$ &$0.024$  \\
$455$ &SFGmain &$3$:$32$:$01.311$ &$-27$:$42$:$43.869$ &$2.300$ &$23.619 \pm 0.029$ &$24.360 \pm 0.059$ &$0.36$ &$0.31$ &$0.029$ &$0.005$  \\
$462$ &SFGmain &$3$:$32$:$01.506$ &$-27$:$44$:$22.597$ &$2.512$ &$24.234 \pm 0.068$ &$25.206 \pm 0.120$ &$0.22$ &$0.15$ &$0.024$ &$0.009$  \\
$479$ &SFGmain &$3$:$32$:$02.063$ &$-27$:$45$:$13.900$ &$2.313$ &$23.729 \pm 0.052$ &$23.653 \pm 0.026$ &$0.14$ &$0.07$ &$0.047$ &$0.009$  \\
$482$ &SFGhighz &$3$:$32$:$02.163$ &$-27$:$42$:$30.408$ &$3.263$ &$24.108 \pm 0.046$ &$24.971 \pm 0.103$ &$0.29$ &$0.24$ &$0.022$ &$0.011$  \\
\enddata

\tablenotetext{a}{Position of WFC3 $H_{160}$-band centroid}
\tablenotetext{b}{Half-light radius computed within $1\farcs2$ fixed aperture before convolution}
\tablenotetext{c}{Internal color dispersion between F160W and F775W images}
\tablenotetext{d}{Uncertainty on internal color dispersion based on Monte Carlo simulations (see Section~\ref{subsec:Simulations})}

\end{deluxetable}

\end{document}